\documentclass[aip,jcp,reprint,amsmath,amssymb]{revtex4-2}
\usepackage{graphicx}
\usepackage{natmove}

\newcommand{\vv}[1]{\mathbf{#1}}

\begin{document}

\title{Simulating hydrodynamic interactions in colloidal suspensions using multiparticle collision dynamics with rigid-body constraints}

\author{Michaela Bush}
\affiliation{Department of Chemical Engineering, Auburn University, Auburn, AL 36849, USA}

\author{Jeremy C. Palmer}
\affiliation{Department of Chemical and Biomolecular Engineering, University of Houston, Houston, TX 77204, USA}

\author{Michael P. Howard}
\email{mphoward@auburn.edu}
\affiliation{Department of Chemical Engineering, Auburn University, Auburn, AL 36849, USA}

\begin{abstract}
We develop a method for simulating colloidal suspensions using multiparticle collision dynamics (MPCD) with a discrete particle model represented as a rigid body. The key steps for incorporating the rigid-body constraints are to thermalize the velocities of the discrete sites before they participate in the MPCD collision step, then transfer momentum from the sites to the rigid body. We demonstrate that the rigid-body model produces the expected statistics for a single spherical particle and the same transport properties for a hard-sphere colloidal suspension as an equivalent model using harmonic bonds to maintain the site geometry. Importantly, the rigid-body model has less computational overhead and permits a larger simulation timestep than the harmonic-bond model, leading to a nearly order of magnitude speedup in benchmark simulations of hard-sphere colloidal suspensions. Our method is compatible with arbitrary discretization, so it enables more efficient MPCD simulations of suspensions of colloidal particles with complex shapes.
\end{abstract}

\maketitle

\section{Introduction}
The motion of colloidal particles suspended in a fluid is important in many natural phenomena and technological applications, ranging from diffusiophoresis in cells \cite{sear_diffusiophoresis_2019} to the directed assembly of nanocrystals into optoelectronic devices \cite{boles_self-assembly_2016}. Computer simulations have proven to be useful tools for investigating the dynamics and emergent behavior of colloidal suspensions \cite{bolintineanu_particle_2014, howard_modeling_2019,delmotte_modeling_2025}. However, simulating colloidal suspensions is challenging because colloidal particles are nanometer to micrometer in size and hence evolve on spatiotemporal scales far longer than those of the solvent \cite{padding_hydrodynamic_2006}, making it often impractical to fully resolve both. To address this challenge, several mesoscale simulation methods have been developed to capture the effects of the solvent, at reduced computational cost, as hydrodynamic interactions (HI) that transfer momentum between colloidal particles. Prominent examples include implicit-solvent models such as Brownian \cite{allen_computer_2017} and Stokesian\cite{brady_stokesian_1988} dynamics, lattice-based models such as the lattice Boltzmann method \cite{chen_lattice_1998, van_den_akker_lattice_2018}, and particle-based models such as dissipative particle dynamics (DPD) \cite{hoogerbrugge_simulating_1992, espanol_perspective_2017} and multiparticle collision dynamics (MPCD) \cite{malevanets_mesoscopic_1999}.

MPCD is a particularly versatile mesoscale method because it uses a highly simplified solvent model that naturally includes thermal fluctuations, readily couples to solid boundaries, and is compatible with a variety of coarse-grained models for soft materials \cite{kapral_multiparticle_2008, gompper_multi-particle_2009, howard_modeling_2019}. Instead of interacting through pairwise conservative forces, MPCD solvent particles are periodically grouped into spatially localized cells and stochastically exchange momentum with particles in the same cell through a prescribed collision scheme. All collision schemes must conserve mass and linear momentum in a cell, but they can be formulated to conserve other quantities such as angular momentum \cite{noguchi_particle-based_2007, noguchi_transport_2008, gotze_relevance_2007, yang_effect_2015, theers_effects_2014} or kinetic energy\cite{malevanets_mesoscopic_1999}. Constant temperature can also be achieved through the collision rule itself \cite{allahyarov_mesoscopic_2002, noguchi_particle-based_2007} or by adding a thermostat \cite{huang_cell-level_2010, huang_thermostat_2015}.

Colloidal suspensions are simulated in MPCD with a hybrid scheme that uses conventional molecular dynamics (MD) methods for the colloidal particles with coupling to the MPCD solvent.\cite{malevanets_dynamics_2000} One family of coupling approaches treats the colloidal particles as solid objects that reflect solvent particles using pair potentials \cite{padding_hydrodynamic_2006} or bounce-back schemes \cite{padding_stick_2005,whitmer_fluidsolid_2010}. However, these approaches require calculation of pairwise distances between the solvent and colloidal particles, resulting in considerable computational overhead. Care must also be taken to mitigate artifacts associated with the excluded volume of the colloidal particle \cite{padding_hydrodynamic_2006} and collision cells intersected by the surface of the colloidal particle \cite{whitmer_fluidsolid_2010}. These considerations become increasingly challenging when the colloidal particles have complex shapes. 

Many of these difficulties can be circumvented using an alternative coupling approach based on a discrete particle model \cite{poblete_hydrodynamics_2014}---also referred to as a composite bead \cite{swan_rapid_2016}, raspberry \cite{lobaskin_new_2004}, or multiblob \cite{balboa_usabiaga_hydrodynamics_2016} model---that represents the surface of the colloidal particle using a collection of sites. The surface sites exchange momentum with the solvent through the cell-based collisions. The solvent passes freely through the colloidal particle between collisions so no pairwise distance calculations are required. This approach is hence straightforward to implement and to adapt to colloidal particles with complex shapes. Discrete particle models have been shown to successfully capture short-range HI between 2 spherical colloidal particles \cite{peng_multiparticle_2024}, as well as the long-time diffusion and sedimentation of colloidal suspensions of monodisperse spherical particles \cite{wani_diffusion_2022}, bidisperse spherical particles \cite{howard_transport_2026}, and shape-anisotropic particles \cite{wani_mesoscale_2024}. 

An important technical consideration for a discrete particle model is that the surface sites must faithfully represent the colloidal particle during the simulation. Typically in MPCD, a network of stiff harmonic bonds is used to maintain the particle shape; however, the number of bonds required may introduce sizable computational overhead. A large spring constant is also required to minimize shape fluctuations, and a small MD timestep is needed to properly integrate the high-frequency vibrational modes associated with these bonds, further increasing computational cost. In principle, these limitations can be overcome by treating the surface sites as a true rigid body that translates and rotates with the colloidal particle rather than having their own degrees of freedom. Indeed, rigid constraints are widely used in MD simulations to eliminate fast dynamic modes and permit a larger integration timestep \cite{allen_computer_2017}, and they have been successfully used to perform implicit-solvent \cite{delong_brownian_2015}, lattice-Boltzmann\cite{mackay_coupling_2013}, and DPD \cite{barriuso_gutierrez_simulating_2022,g_sedimentation_2025} simulations of colloidal suspensions using discrete particle models. However, to our knowledge, rigid-body constraints have not yet been adapted to MPCD.

In this article, we investigate the accuracy and performance of MPCD simulations of colloidal suspensions using a discrete particle model with rigid-body constraints. We show that naively coupling surface sites with rigid constraints to the MPCD solvent through momentum-conserving collisions leads to incorrect partitioning of energy because the surface sites do not have independent degrees of freedom. We propose a scheme to resolve this issue through a random thermalization of momenta for the surface sites prior to the collision, which is inspired by the physics of harmonic-bond models used previously \cite{poblete_hydrodynamics_2014, wani_diffusion_2022, peng_multiparticle_2024}. We show that our method yields transport properties for colloidal suspensions of monodisperse spherical particles that are consistent with both theoretical expectations and previous MPCD simulations using a harmonic-bond model. Importantly, the rigid-body constraints both reduce the computational cost associated with a single timestep and allow a larger MD integration timestep compared to the harmonic bonds, resulting in a considerable computational speedup in the simulations over the current state of the art.

The rest of this article is organized as follows. Section \ref{sec:model} describes the models used for the solvent and the colloidal particles, focusing on the implementation of rigid-body constraints for the discrete particle model. Section \ref{sec:results} compares results of simulations using the rigid-body model with expected energy and velocity distributions, velocity autocorrelation functions, and suspension transport properties. It also reports a benchmark for the rigid-body model compared to an equivalent harmonic-bond model. Section \ref{sec:conclusions} summarizes these findings.

\section{Model}
\label{sec:model}
All quantities will be reported in a system of units where $\ell$ is the unit of length, $m$ is the unit of mass, and $\varepsilon$ is the unit of energy. The unit of temperature is $\varepsilon/k_{\rm B}$, where $k_{\rm B}$ is the Boltzmann constant, and the unit of time is $\tau = \sqrt{m\ell^2/\varepsilon}$.

\subsection{Solvent}
We simulated the solvent using MPCD, in which the dynamics of solvent particles are governed by alternating streaming and collision steps \cite{kapral_multiparticle_2008, gompper_multi-particle_2009, howard_modeling_2019}. In a streaming step, the positions and velocities of the solvent particles evolve over a chosen time interval according to Newton's equations of motion. Unlike in conventional MD, there are no pairwise interactions between the solvent particles, so this time integration is often simple to perform. In a collision step, the solvent particles are sorted into cells and exchange momentum with other particles in their cell according to the chosen collision rule. For example, in the stochastic rotation dynamics (SRD) collision rule, the velocity of a solvent particle relative to the center-of-mass velocity of its cell is rotated using an operator randomly generated for each cell \cite{allahyarov_mesoscopic_2002}. This version of the SRD collision rule conserves both energy and linear momentum within a cell but not angular momentum. It can be modified to enforce a temperature $T$ through a cell-level thermostat that randomly rescales the cell's kinetic energy according to the expected distribution \cite{huang_cell-level_2010, huang_thermostat_2015}, as well as to conserve angular momentum \cite{noguchi_transport_2008}.

We used the SRD collision rule without angular momentum conservation, a thermostat set to $T=1\,\varepsilon/k_{\rm B}$, and cubic collision cells with edge length $1\,\ell$. The solvent particles had mass $1\,m$, and the solvent mass density was $\rho_0 = 5\,m/\ell^3$ (5 particles per cell). Collisions were performed every $0.1\tau$ using a fixed rotation angle of $130^{\circ}$ about an axis randomly selected from the unit sphere. The collision cells were randomly shifted along the Cartesian axes by an amount drawn uniformly in $[-\ell/2,\ell/2]$ before each collision to ensure Galilean invariance. \cite{ihle_stochastic_2003} These common choices give a solvent dynamic viscosity $\eta_0 = 3.96\,\varepsilon \tau/\ell^3$ \cite{statt_unexpected_2019,kikuchi_transport_2003,tuzel_transport_2003}.

\subsection{Colloidal particles}
We simulated spherical colloidal particles with diameter $d=6\,\ell$. The colloidal particles interacted with each other as nearly hard spheres through the core-shifted Weeks--Chandler--Andersen potential:
\begin{equation}
u(r)= \begin{cases} 
    \displaystyle 4\varepsilon\left[ \left( \frac{\sigma}{r - \Delta} \right)^{12} -\left( \frac{\sigma}{r - \Delta}  \right)^{6} \right]+1, & r\leq \Delta+2^{1/6}\sigma \\
    0, & \text{otherwise}
\end{cases},
\label{eq:WCA_potential}
\end{equation}
where $r$ is the distance between the centers of the particles, $\sigma=1\ell$ sets the steepness of the repulsion, and $\Delta = d-\sigma=5\,\ell$ is the shift factor to account for particle size. The colloidal particles were coupled to the solvent using a discrete particle model constructed by iteratively subdividing the faces of a regular icosahedron into equilateral triangles, then rescaling the resulting vertices to lie on the surface of the sphere \cite{wani_diffusion_2022, howard_transport_2026}. We performed this subdivision step twice to create 162 surface sites, giving a surface density of approximately $1.43\,\ell^{-2}$ This surface density was shown to produce good results for many dynamic properties of colloidal particles with the same diameter in an MPCD solvent \cite{peng_multiparticle_2024}. An additional site was placed at the center of the sphere solely for performing the excluded volume calculations. The mass of each site was set to $5\,m$ based on our chosen solvent density and prior work \cite{poblete_hydrodynamics_2014, wani_diffusion_2022, wani_mesoscale_2024, howard_transport_2026}.

We will compare two different strategies for evolving the dynamics of the colloidal particles while ensuring the sites maintain their desired geometry: a conventional strategy based on harmonic bonds and a new strategy based on rigid-body constraints. We describe the details of these two strategies next.

\subsubsection{Harmonic-bond model}
Following established practice \cite{poblete_hydrodynamics_2014}, a network of harmonic bonds was used to maintain a nearly rigid configuration of surface sites. Each surface site was bonded to its three nearest-neighbor surface sites and to the central site using a harmonic potential:
\begin{equation}
u_{\rm b}(r)=\frac{k_{\rm b}}{2}(r-r_{\rm b})^2
\label{eq:harmonic}
\end{equation}
where $r$ is the distance between sites, $k_{\rm b}$ is the spring constant, and $r_{\rm b}$ is the unstretched bond length. We calculated $r_{\rm b}$ for all the bonds in our model, and we rounded to three decimal places to reduce the number of different bond lengths considered. A large spring constant is needed to maintain rigidity, so we used $k_{\rm b}=5000\,\varepsilon/\ell^2$ based on prior work \cite{wani_diffusion_2022, wani_mesoscale_2024, howard_transport_2026}. A range of values has been used for spring constants in the literature, but they are of similar magnitude \cite{hu_modelling_2015,das_colloidal_2019,clopes_flagellar_2021,mahdiyehmousavi_wall_2020, rusenargun_influence_2023,poblete_hydrodynamics_2014,peng_multiparticle_2024,kobayashi_structure_2020,yokoyama_aggregation_2023}. We will refer to this strategy as the harmonic-bond model.

The surface sites for the harmonic-bond model participated in the collision step, and they were treated identically to the solvent particles except for their heavier mass. Between collisions, the sites were acted on by the forces resulting from Eqs.~\eqref{eq:WCA_potential} and \eqref{eq:harmonic}, and their positions and velocities were integrated using the velocity Verlet algorithm with timestep $\Delta t$. One of us recently characterized the transport properties of suspensions of spherical colloidal particles as a function of volume fraction using this approach with the same model parameters and timestep $\Delta t = 0.002\,\tau$ \cite{howard_transport_2026}. We will hence primarily cite these results here rather than repeat them, but we performed some simulations of the harmonic-bond model using different values of the timestep. 

\subsubsection{Rigid-body model}
A practical challenge of the harmonic-bond model is that $\Delta t$ must be sufficiently small to faithfully integrate the dynamics of the sites, including the fast vibrations associated with the bonds; however, these vibrations are not of physical interest because the bonds are an artificial construction. We are instead interested in the collective translation and rotation of the sites, which directly represent the motion of the colloidal particle and typically occur over longer time scales. A rigid body is a potentially better representation of this motion.

Specifically, we consider a rigid body consisting of $N$ sites, and we denote the position of site $j$ as $\vv{r}_j$, the velocity of site $j$ as $\vv{v}_j$, and the mass of site $j$ as $m_j$. The position of the rigid body is represented by the center of mass $\vv{R}$ calculated from the sites, while its orientation is represented by the rotation $\boldsymbol{\Omega}$ of the sites about the center of mass. These coordinates have associated linear momentum $\vv{P}$ and angular momentum $\vv{L}$, respectively,
\begin{align}
\mathbf{P} &= \sum_{i=1}^N m_j \vv{v}_j \label{eq:linmom}\\
\mathbf{L} &= \sum_{i=1}^N \vv{d}_j \times (m_j \vv{v}_j) \label{eq:angmom},
\end{align}
where $\mathbf{d}_j = \vv{r}_j-\vv{R}$ is the displacement vector of site $j$ from the center of mass. Rigidity means that all sites translate and rotate together, which requires that
\begin{equation}
\vv{v}_j = \vv{V} + \boldsymbol{\omega} \times \vv{d}_j,
\label{eq:rigid}
\end{equation}
where $\vv{V} = \vv{P}/M$ and $\boldsymbol{\omega} = \vv{I}^{-1} \cdot \vv{L}$ are the linear and angular velocities of the body calculated using its total mass $M$ and moment of inertia tensor $\vv{I}$. The body is acted on by forces and torques that can be expressed as functions of $\vv{R}$ and $\boldsymbol{\Omega}$ or calculated from the forces acting on the sites \cite{allen_computer_2017}, but the sites cannot deform as a result of these interactions. The equations of motion for rigid bodies can be formulated and numerically integrated using various schemes. Here, we used the quaternion-based scheme of Miller et. al.\cite{miller_symplectic_2002} with velocity Verlet integration that is available in widely-used molecular simulation software packages such as HOOMD-blue\cite{anderson_hoomd-blue_2020,nguyen_rigid_2011,glaser_pressure_2020} and LAMMPS\cite{thompson_lammps_2022}. We will refer to this strategy as the rigid-body model.

The surface sites for the rigid-body model also participated in the collision step. The velocities of the sites before the collision were consistent with rigid motion [Eq.~\eqref{eq:rigid}], but the velocities immediately after the collision were not, in general, as a result of the momentum exchange. Naively, the net change in the linear momentum $\Delta\vv{P}$ and angular momentum $\Delta \vv{L}$ of the sites could be calculated using Eqs.~\eqref{eq:linmom} and \eqref{eq:angmom} with the site velocities before and after the collision, then $\Delta\vv{P}$ and $\Delta\vv{L}$ could be used to update $\vv{P}$ and $\vv{L}$, and finally, the site velocities could be reset according to Eq.~\eqref{eq:rigid}. This procedure conserves the linear and angular momentum of the body and satisfies the rigid-body constraints, but it does not generally conserve energy. Indeed, we implemented this naive scheme and found in preliminary simulations of a colloidal suspension that the total energy decreased toward zero without a thermostat and even with a thermostat, the correct average kinetic energy was not achieved (Fig.~S1).

Qualitatively, this issue arises because the surface sites are treated as having independent degrees of freedom during the collision, but the rigid-body constraints remove dynamic modes that are not associated with translation or rotation of the body \cite{palmer_comment_2018}. This issue is not present for the harmonic-bond model because, although nearly rigid, all sites can still move independently. The sites in the harmonic-bond model should quickly exchange energy with each other and also come to thermal equilibrium with the solvent through the collision step. Hence, we expect their degrees of freedom to have roughly their equilibrium distribution at constant temperature.

\begin{figure}
  \centering
  \includegraphics{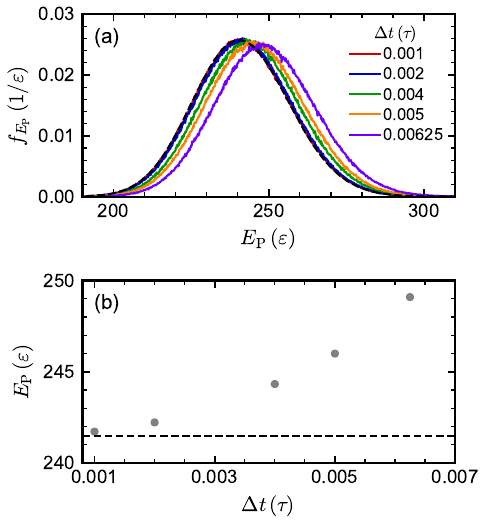}
  \caption{(a) Probability density $f_{E_{\rm P}}$ for the potential energy $E_{\rm P}$ of a single particle represented using the harmonic-bond model and different integration timesteps $\Delta t$. The particle was simulated in a cubic box with edge length $30\,\ell$ and, after a warm up period of $10^{3}\tau$, the potential energy was sampled every $1\,\tau$ for $5 \times 10^5\,\tau$. The probability density was calculated using a histogram with bin width $0.15\,\varepsilon$. (b) Average potential energy $\langle E_{\rm P} \rangle$ as a function of $\Delta t$. The black dashed lines are expectations from Eq.~\eqref{eq:harmonicpdf} with $n = 3N-6$.}
  \label{fig:spring_potential}
\end{figure}

To demonstrate, we computed the potential energy $E_{\rm P}$ of a single colloidal particle in the MPCD solvent using the harmonic-bond model, finding that its average value converged to a finite nonzero value. This value can be interpreted using a normal mode analysis: for a sufficiently large spring constant, the potential energy can be series expanded to second order in the displacements of the particles from their desired positions, giving $3N-6$ vibrational modes that contribute harmonically to the potential energy. The expected probability density for the energy $E$ for $n$ harmonic degrees of freedom at constant temperature is
\begin{equation}
f_E(E, n) = \frac{\beta}{\Gamma(n/2)} (\beta E)^{n/2-1} e^{-\beta E},
\label{eq:harmonicpdf}
\end{equation}
where $\beta=1/(k_{\rm B}T)$. The expected value of the energy is $\langle E \rangle = n k_{\rm B}T/2$. We confirmed that the measured $E_{\rm P}$ converged toward this distribution and expected value as the timestep became sufficiently small (Fig.~\ref{fig:spring_potential}). Note that these results also highlight how small a timestep is required for faithful integration.

Based on this analysis, we proposed to reconsider the rigid-body model as a harmonic-bond model with an infinitely large spring constant for purposes of the collision step. Within this picture, each component of the site velocities should be Maxwell--Boltzmann distributed, giving $3N$ degrees of freedom. These degrees of freedom can be reexpressed as 3 degrees of freedom associated with the net linear velocity $\vv{V}$, 3 degrees of freedom associated with the net angular velocity $\boldsymbol{\omega}$, and the remaining $3N-6$ degrees of freedom associated with internal motion. Before each collision step, each site $j$ in the rigid body was given a random velocity $\vv{\tilde v}_j$ drawn from the Maxwell--Boltzmann distribution at the temperature of the solvent $T$. The linear velocity $\vv{\tilde V}$ and angular velocity $\boldsymbol{\tilde\omega}$ associated with the random velocities were calculated, then the random velocities were added to the current site velocities with this motion removed so that $\vv{V}$ and $\boldsymbol{\omega}$ did not change:
\begin{equation}
\vv{v}_j \gets \vv{V} - \vv{\tilde V} + (\boldsymbol{\omega} - \boldsymbol{\tilde \omega}) \times \vv{d}_j + \vv{\tilde{v}}_j.
\end{equation}
This thermalization procedure hence conserved the linear and angular momentum of the rigid body, but it intentionally violated the rigid-body constraint so that each site velocity was independent for the purposes of the collision. The surface sites then participated in the collision, the net change in linear and angular momentum of the body was transferred, and the velocities of all sites were reset using Eq.~\eqref{eq:rigid}. Between collisions, the sites were acted on by only Eq.~\eqref{eq:WCA_potential}, and their positions were integrated using the velocity Verlet algorithm \cite{miller_symplectic_2002}.

We note that a similar thermalization strategy has been used in MPCD when the colloidal particles are coupled to the solvent during the streaming step using specular reflection \cite{lamura_multi-particle_2001}. In that scheme, virtual particles needed to be added to the colloidal particle to address issues with underfilled collision cells, and the momentum change of the virtual particles was transferred to the colloidal particle \cite{whitmer_fluidsolid_2010}. The virtual particles had their velocities drawn from a Maxwell--Boltzmann distribution whose average velocity was the velocity of the colloidal surface, but their mean linear and angular velocities were not zeroed like in our proposed strategy. Some issues with the temperature of the colloidal particle were reported \cite{whitmer_fluidsolid_2010}, emphasizing the sensitivity of thermalization methods to implementation details.

We also note that our thermalization scheme requires setting a temperature, so it is most compatible with a thermostatted collision method \cite{huang_thermostat_2015,allahyarov_mesoscopic_2002}. It may be possible to generalize the scheme to make the simulation energy-conserving by treating the kinetic energy of the surface sites as a random variable in the microcanonical ensemble associated with the total energy of the vibrational degrees of freedom. This extension may be an interesting avenue to explore in future work. 

\section{Results and discussion}
\label{sec:results}
We implemented our rigid-body model in a development branch of HOOMD-blue \cite{anderson_hoomd-blue_2020, howard_efficient_2018,howard_quantized_2019} based on version 5.4.0, which we used to generate the results in this article. The features in the development branch were publicly released in version 6.0.0. The integration timestep for the rigid-body model was $\Delta t =0.1\tau$ unless otherwise specified.

\subsection{Single particle validation}
We first verified that the rigid-body model produced correct statistics for the linear and angular velocities. We simulated a single particle in a cubic box with edge length $90\,\ell$ and periodic boundary conditions. After an equilibration period of $10^3\,\tau$, the linear velocity $\vv{V}$ and the angular velocity $\boldsymbol{\omega}$ of the particle were recorded every $1\,\tau$ for $5 \times 10^6\,\tau$. We also recorded the associated translational $E_{\rm T} = M |\vv{V}|^2/2$ and rotational $E_{\rm R} = \boldsymbol{\omega}\cdot\vv{I}\cdot\boldsymbol{\omega}/2$ kinetic energies. The recorded angular velocities were rotated to a body-fixed reference frame in which the moment of inertia tensor was diagonal for some of the analysis, and quantities in this reference frame are denoted by a hat ($\boldsymbol{\hat{\omega}}$ and $\vv{\hat{I}}$). This procedure was repeated 10 times.

\begin{figure}
  \centering
  \includegraphics{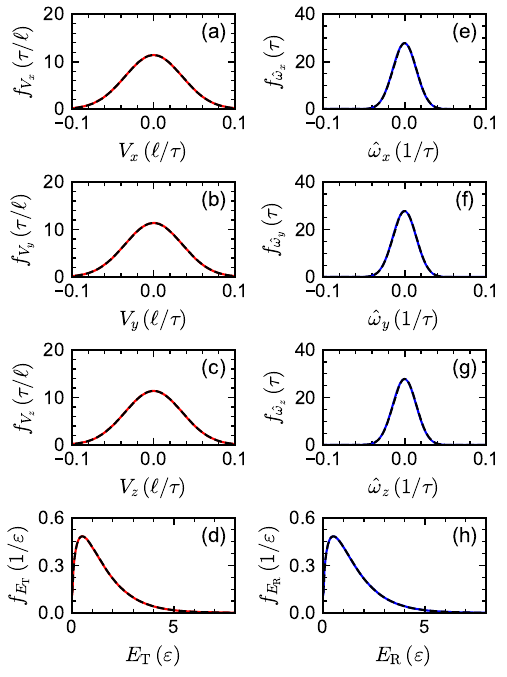}
  \caption{Probability densities for the components of the (a)--(c) linear velocity $\vv{V}$, (d) translational kinetic energy $E_{\rm T}$, (e)--(g) body-fixed angular velocity $\boldsymbol{\hat{\omega}}$, and (h) rotational kinetic energy $E_{\rm R}$. The black dashed lines are the expected distributions based on Eqs.~\eqref{eq:harmonicpdf}, \eqref{eq:mbtrans}, and \eqref{eq:mbrot}.}
  \label{fig:energy_distribution}
\end{figure}

The components of both $\vv{V}$ and $\boldsymbol{\hat{\omega}}$ are expected to have Gaussian-distributed probability densities,
\begin{align}
f_{V_\alpha}(V_\alpha) &= \sqrt{\frac{M\beta}{2\pi}} e^{-\beta MV_\alpha^2/2} \label{eq:mbtrans} \\
f_{\hat{\omega}_\alpha}(\hat{\omega}_\alpha) &= \sqrt{\frac{\hat{I}_{\alpha\alpha}\beta}{2\pi}} e^{-\beta \hat{I}_{\alpha\alpha} \hat{\omega}_\alpha^2/2} \label{eq:mbrot}
\end{align}
respectively, where $\alpha$ denotes an index of the respective vector or tensor. We calculated the probability density of these velocity components in our simulations using histograms with bin widths $4\times10^{-4}\,\ell/\tau$ and $4\times10^{-4}\,\tau^{-1}$ for $\vv{V}$ and $\boldsymbol{\hat\omega}$, respectively, finding excellent agreement (Fig.~\ref{fig:energy_distribution}). We also computed the probability density of the energies $E_{\rm T}$ and $E_{\rm R}$ using a histogram with bin width $0.025\,\varepsilon$, which should follow Eq.~\eqref{eq:harmonicpdf} with $n=3$; we again found excellent agreement. Further, the average translational kinetic energy was $\langle E_{\rm T} \rangle = 1.4999 \pm 0.0003\,\varepsilon$ and the average rotational kinetic energy was $\langle E_{\rm R} \rangle = 1.50015 \pm 0.00009\,\varepsilon$, both of which are in good agreement with the expected value of $1.5\,\varepsilon$.

To more rigorously check for agreement between the simulations and theoretical expectations, we also applied the statistical validation tests developed by Shirts and coworkers. \cite{shirts_simple_2013, merz_testing_2018} These tests check against the null hypotheses that the total kinetic energy is distributed according to Eq.~\eqref{eq:harmonicpdf} and that the kinetic energy associated with every subset of the degrees of freedom (e.g., one component of velocity) is also distributed according to Eq.~\eqref{eq:harmonicpdf}.\cite{merz_testing_2018} Both tests were performed on the translational and rotational degrees of freedom separately using every fifth sample from a single simulation, and the null hypotheses could not be rejected. This analysis confirms that our proposed thermalization method for the rigid-body model produced distributions of linear and angular velocities consistent with Maxwell--Boltzmann statistics. 

We then proceeded to analyze the autocorrelation of the linear and angular velocities. These autocorrelation functions are connected to the translational and rotational diffusion of the particle by Green--Kubo relations.\cite{allen_computer_2017} The linear velocity autocorrelation function $C_{\rm T}(t) = \langle \vv{V}(0) \cdot \vv{V}(t) \rangle$ and angular velocity autocorrelation function $C_{\rm R}(t) = \langle \boldsymbol{\omega}(0) \cdot \boldsymbol{\omega}(t) \rangle$ were computed for times $t \le 50\,\tau$ using all recorded values and every value as a new time origin. 

An expression for $C_{\rm T}$ has been derived for a neutrally-buoyant sphere with no-slip boundary conditions based on a frequency-dependent translational friction coefficient $\gamma_{\rm T}$ \cite{montgomery_molecular_1977}. It is defined as an inverse Fourier transform,
\begin{equation}
C_{\rm T}\left(t\right)=\frac{3k_BT}{2\pi}\int_{-\infty}^{\infty} \frac{e^{-i \omega t}}{\gamma_{\rm T}-i \omega M_{\rm S}} {\rm d}\omega,
\label{eq:VACF_fourier}
\end{equation}
where $i$ is the imaginary unit, $M_{\rm S} = 4\pi R^3\rho_0/3$ is the mass of a neutrally-buoyant sphere with radius $R=d/2$, and $\omega$ in this context refers to frequency.\cite{poblete_hydrodynamics_2014,felderhof_backtracking_2005} The expression for the translational friction coefficient is \cite{felderhof_backtracking_2005,zwanzig_hydrodynamic_1970, metiu_hydrodynamic_1977}
\begin{equation}
\gamma_{\rm T} =\frac{4\pi}{3}\eta_0\ Rx^2\frac{(1+x)(9-9i\zeta-2\zeta^2)+x^2(1-i\zeta)}{2x^2(1-i\zeta)-(1+x)\zeta^2-x^2\zeta^2},
\label{eq:translational_friction}
\end{equation}
where $x=R\sqrt{-i\omega/\nu_0}$, $\zeta=s R/\sqrt{c_0^2-i\omega(4\eta_0/3 + \eta_{0, {\rm b}})/\rho_0}$, $c_0$ is the speed of sound in the solvent, $\nu_0 = \eta_0/\rho_0$ is the kinematic viscosity of the solvent, and $\eta_{0,{\rm b}}$ is the bulk viscosity of the solvent. For our solvent, the isothermal speed of sound is $c_0 = 1\,\ell/\tau$ \cite{huang_hydrodynamic_2012}, the kinematic viscosity is $\nu_0 = 0.792\,\ell^2/\tau$, and the bulk viscosity is $\eta_{0,{\rm b}} = 1.22\,\varepsilon \tau/\ell^3$ \cite{theers_bulk_2015}. We evaluated Eq.~\eqref{eq:VACF_fourier} using the trapezoid rule with an integration step of $0.001\,\tau^{-1}$ and bounds of $\pm 1000\,\tau^{-1}$.

\begin{figure}
  \centering
  \includegraphics{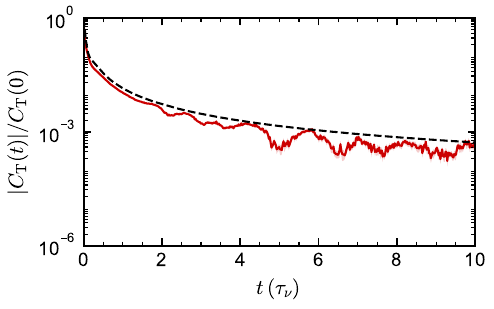}
  \caption{Linear velocity autocorrelation function $C_{\rm T}$ of rigid-body model (red solid line) compared to the theoretical expectation for a neutrally-buoyant sphere (Eq.~\eqref{eq:VACF_fourier}, dashed black line). The time $t$ is given in units of the characteristic viscous time scale $\tau_\nu = d^2/\nu_0 = 45.5\,\tau$.}
  \label{fig:VACF}
\end{figure}

The initial value of the linear velocity autocorrelation function, $C_{\rm T}(0)$, is expected to be different between the simulations and Eq.~\eqref{eq:VACF_fourier} because the mass of the discrete particle model $M$ is different from the mass of a neutrally buoyant sphere $M_{\rm S}$. Hence, we normalized $C_{\rm T}$ by $C_{\rm T}(0)$ to compare the simulations and Eq.~\eqref{eq:VACF_fourier}, finding good agreement in their decay behavior (Fig.~\ref{fig:VACF}). At short times, $C_{\rm T}$ decayed at a similar rate in the simulations and according to Eq.~\eqref{eq:VACF_fourier}. At longer times, $C_{\rm T}$ for the simulations continued to decay at roughly the same rate as Eq.~\eqref{eq:VACF_fourier} but with some undulations. These undulations are expected based on prior work using discrete particle models in MPCD \cite{poblete_hydrodynamics_2014}.

\begin{figure}
  \centering
  \includegraphics{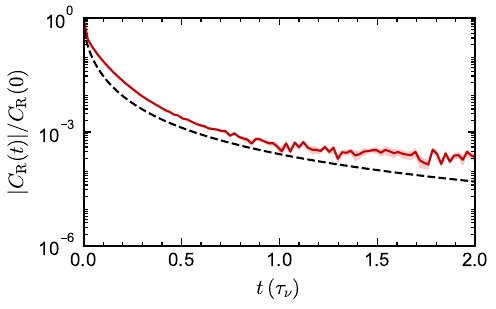}
  \caption{Angular velocity autocorrelation $C_{\rm R}$ of rigid-body model (red solid line) compared to the theoretical expectation for a neutrally-buoyant sphere (Eq.~\eqref{eq:AVACF_laplace}, dashed black line).}
  \label{fig:AVACF}
\end{figure}

An expression for $C_{\rm R}$ has also been derived for a neutrally-buoyant sphere with no-slip boundary conditions based on a frequency-dependent rotational friction coefficient $\gamma_{\rm R}$ \cite{montgomery_molecular_1977}. It is defined as an inverse Laplace transform,
\begin{equation}
C_{\rm R}(t)=\frac{3k_{\rm B} T}{I_{\rm S}}\frac{1}{2\pi i}\int_{s_0-i\infty}^{s_0+i\infty} \frac{e^{st}}{s+\gamma_{\rm R}/I_{\rm S}}\,{\rm d}s,
\label{eq:AVACF_laplace}
\end{equation}
where $s_0 > 0$ is chosen to successfully perform the integration and $I_{\rm S} = 2 M_{\rm S} R^2 / 5$ is the moment of inertia of a solid sphere. The rotational friction coefficient is \cite{montgomery_hydrodynamic_1977, montgomery_molecular_1977}
\begin{equation}
\gamma_{\rm R} = 8\pi\eta R^3\left[1+\frac{R^2s/\nu_0}{3(1+\sqrt{R^2 s/\nu_0})}\right].
\end{equation}
We evaluated Eq.~\eqref{eq:AVACF_laplace} using mpmath (version 1.3.0) \cite{mpmath}. Like $C_{\rm T}$, the values of $C_{\rm R}(0)$ were different between the simulation and Eq. \eqref{eq:AVACF_laplace} due to the difference between the moment of inertia of the sites and $I_{\rm S}$, so we normalized by $C_{\rm R}(0)$ (Fig.~\ref{fig:AVACF}). We again found good agreement between the simulations and theoretical exceptions [Eq.~\eqref{eq:AVACF_laplace}], with $C_{\rm R}$ decaying somewhat slower in the simulations than predicted. All together, this analysis further confirms the validity of the thermalization method.

\subsection{Suspension validation}
We next characterized the properties of colloidal suspensions simulated using the rigid-body model. The particle volume fraction $\phi$ was varied between 0.01 and 0.4 in a cubic box with edge length $L=120\,\ell$ and periodic boundary conditions. Five equilibrated particle configurations were taken from Ref.~\citenum{howard_transport_2026} and used to perform independent simulations of the shear viscosity, long-time self-diffusion coefficient, and sedimentation velocity. Error bars were estimated from the standard error between the independent simulations. The simulation protocols are essentially the same as in Ref.~\citenum{howard_transport_2026}, but they are summarized here for completeness.

The suspension shear viscosity $\eta$ was characterized using reverse nonequilibrium simulations \cite{muller-plathe_reversing_1999,tenney_limitations_2010}. For these simulations, the cubic simulation box was duplicated in the $y$ direction to give a box length $L_y = 2L=240\,\ell$. A bidirectional linear shear flow in the $x$-direction was generated by exchanging momentum between solvent particles in two planar slabs with normal in the $y$ direction and width $\Delta y=1\,\ell$. Specifically, the lower slab was located at $-L_y/2 \leq y \leq -L_y/2 + \Delta y$, and the upper slab was located at $0 \leq y \leq \Delta y$. Every $0.1\,\tau$, the solvent particles in the lower slab whose velocities in the $x$ direction were closest to a target value $0.5\,\ell/\tau$ and in the upper slab whose velocities in the $x$ direction were closest to $-0.5\,\ell/\tau$ were identified in sorted order. Up to the first 100 pairs of sorted solvent particles then exchanged the $x$-component of their velocities. After a warmup simulation of $2 \times 10^4\,\tau$, the cumulative momentum exchanged between the slabs $p_x(t)$ was sampled every $1\,\tau$, and the mass-averaged velocity of the suspension $u_x(y)$ in the $x$ direction as a function of $y$ was sampled every $10\,\tau$ during a $2\times10^4\tau$ production simulation using a histogram with bin width $0.5\,\ell$ (Fig. S2). The average rate of momentum transferred $\dot{p}_x$ was calculated over the entire simulation, and the shear rate $\dot{\gamma}={\rm d}u_x/{\rm d}y$ was calculated from a linear fit of $u_x(y)$ making use of the symmetry of the profile. The suspension viscosity was then calculated as

\begin{equation}
\eta = \frac{1}{2L^2}\frac{\dot{p}_x}{\dot{\gamma}}.
\label{eq:visocity_calc}
\end{equation}

The long-time self-diffusion coefficient $D$ was determined from the particle mean squared displacement in equilibrium simulations. After a warmup period of $10^3\,\tau$, configurations were sampled every $10\,\tau$ for $10^5\tau$, then $D$ was calculated via the Einstein relation\cite{allen_computer_2017}
\begin{equation}
D = \lim_{t \to \infty} \frac{1}{6}\frac{{\rm d}}{{\rm d}t}\left\langle \left| \textbf{R}(t) - \textbf{R}(0) \right|^2 \right\rangle.
\label{eq:diffusion_einstein}
\end{equation}
All configurations were used as time origins and all particles were included to evaluate the average in Eq.~\eqref{eq:diffusion_einstein}. The derivative was evaluated numerically, and the limit was taken by averaging its value for $1000\,\tau \le t \le 2000\,\tau$. Finite-size effects from the periodic boundary conditions \cite{dunweg_molecular_1993,yeh_system-size_2004} were corrected by adding $\Delta D = \xi/(6\pi\eta L)$ to the simulated $D$, where $\xi=2.837297$ and $\eta$ is the measured suspension shear viscosity.

The sedimentation velocity of the particles $U$ was characterized using force-driven nonequilibrium simulations. A force $F = 0.5\,\varepsilon/\ell$ in the $x$ direction was applied at the center site of each particle, and a balancing force was applied to all solvent particles to ensure the entire system was force free. After a warmup period of $10^3\,\tau$, the average velocity of all colloidal particles in the $x$ direction was recorded every $0.2\,\tau$ during a $5 \times 10^4\,\tau$ production simulation, then averaged over the simulation. Finite-size effects from the periodic boundary conditions \cite{mo_method_1994} were corrected by adding $\Delta U = \xi S(0) F/(6\pi\eta L)$, where $S(0)$ is the static structure factor of the suspension that we computed using the isothermal compressibility of the Carnahan--Starling equation of state for hard spheres.\cite{carnahan_equation_1969}

\begin{figure}
  \centering
  \includegraphics{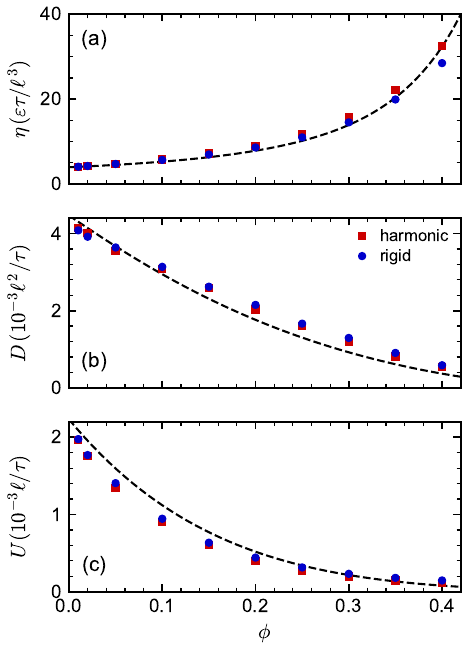}
  \caption{(a) Shear viscosity $\eta$, (b) long-time self-diffusion coefficient $D$, and (c) sedimentation velocity $U$ under applied force $F=0.5\,\varepsilon/\ell$ for hard-sphere suspensions of varying particle volume fraction $\phi$. Simulations with the rigid-body model (blue circles) are compared to simulations from Ref.~\citenum{howard_transport_2026} using the harmonic-bond model (red squares). The simulation results are compared to theoretical predictions of (a) Ref.~\citenum{verberg_viscosity_1997}, (b) Ref.~\citenum{tokuyama_dynamics_1994}, and (c) Ref.~\citenum{wang_short-time_2015-1}, shown as dashed black lines. A summary of these theoretical predictions is also available in Ref.~\citenum{howard_transport_2026}.
}
  \label{fig:transport_coeff}
\end{figure}

We found that the suspension shear viscosity, long-time self-diffusion coefficient, and sedimentation velocity simulated using the rigid-body model (Fig.~\ref{fig:transport_coeff}) were in good agreement with prior results for the harmonic-bond model \cite{howard_transport_2026}. The most noticeable deviation between the two was found for the shear viscosity at the largest particle volume fractions, but the agreement between the two models was generally extremely satisfactory. Further, both the rigid-body and harmonic-bond models were in good agreement with theoretical expectations for hard-sphere suspensions \cite{verberg_viscosity_1997, tokuyama_dynamics_1994, wang_short-time_2015-1}.

\subsection{Benchmark}
Having validated that the rigid-body model produced physics consistent with the harmonic-bond model, we finally assessed whether it achieved our goal to improve simulation performance by removing the fast degrees of freedom associated with bond vibrations. We ran benchmark simulations for hard-sphere colloidal suspensions with varying particle volume fraction using an identical setup as the diffusion simulations on a single NVIDIA A100 GPU on NCSA Delta in September and October 2025 \cite{access}. After a warmup period of $5 \times 10^3\tau$, the average number of timesteps simulated per second over an interval of $10^3\tau$ was recorded 5 times. The median value of these 5 measurements was determined, and the time required to simulate $1\,\tau$ was calculated from it. Last, the average time required to simulate $1\,\tau$ was computed from 5 different simulations.

\begin{figure}
  \centering
  \includegraphics{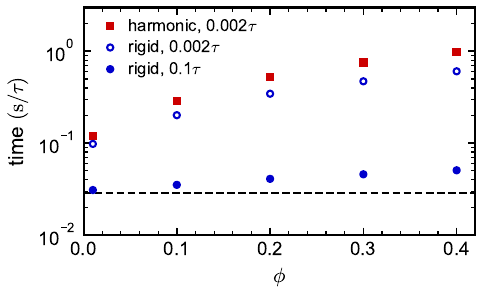}
  \caption{Average time to simulate $1\,\tau$ as a function of volume fraction $\phi$ using the harmonic-bond model with timestep $\Delta t = 0.002\,\tau$, the rigid-body model with timestep $\Delta t = 0.002\,\tau$, and the rigid-body model with timestep $\Delta t = 0.1\,\tau$. The dashed line shows the result of the same benchmark for a pure solvent. The benchmarks were performed on a single NVIDIA A100 GPU on NCSA Delta in September and October 2025 \cite{access}.}
  \label{fig:benchmark}
\end{figure}

We initially performed this benchmark using the same integration timestep $\Delta t = 0.002\,\tau$ for the rigid-body model, which was also used in Ref.~\citenum{howard_transport_2026} for the harmonic-bond model (Fig.~\ref{fig:benchmark}). We found that the rigid-body model required modestly less time than the harmonic-bond model across all volume fractions, with the largest difference being apparent at the largest volume fraction for which there are more colloidal particles. This benchmark shows that the rigid-body model has less computational overhead than the harmonic-bond model during a single timestep, trading off the evaluation of the harmonic bond potentials for the more complex, but less costly, rigid-body dynamics. We then increased the timestep to $\Delta t = 0.1\,\tau$, used for the validation simulations, finding a roughly order of magnitude decrease in the time. Indeed, the computational overhead was only slightly larger than that of the pure solvent for the most dilute volume fraction, and the dependence on volume fraction was also only modest. These observations are consistent with a larger fraction of the simulation time being spent on the MPCD solvent because fewer integration steps are required for the colloidal particles when the bond vibrations do not need to be resolved and a larger timestep can be used. In short, the rigid-body model consistently achieved superior performance to the harmonic-bond model in these benchmarks.

\section{Conclusions}
\label{sec:conclusions}
We have introduced a scheme to accelerate MPCD simulations of colloidal suspensions using discrete particle models with rigid-body constraints. The key steps in the approach are to thermalize the velocities of the sites in the discrete particle model prior to the collision step, then transfer the momentum change of the sites from the collision step to the rigid body. Our approach gives excellent agreement with theoretical expectations for the statistics of the linear and angular velocity of a single spherical particle. It also gives transport properties for hard-sphere colloidal suspensions that are in good agreement with prior work using stiff harmonic bonds to maintain the discrete particle's shape. Importantly, compared to the harmonic-bond model, our approach also decreases the required simulation time by roughly one order of magnitude in benchmarks of hard-sphere colloidal suspensions. Our method can accommodate particles of arbitrary shape (discretization) and is implemented in the MPCD component of HOOMD-blue version 6.0.0. We hence anticipate that our approach will enable new MPCD simulations of colloidal suspensions.

\section*{Supplementary Material}
See the supplementary material for the kinetic energy loss in rigid-body suspension simulations without thermalization and the mass-averaged velocity $u_x$ used to calculate the viscosity at various volume fractions.

\section*{Conflicts of interest}
The authors have no conflicts to disclose.

\section*{Data Availability}
The data that support the findings of this study are available from the authors upon reasonable request.

\section*{Acknowledgments}
This material is based upon work supported by the National Science Foundation under Award Nos.~2310724 and 2310725. JCP gratefully acknowledges support from the Welch Foundation (Grant E-1882). This work used Delta at the National Center for Supercomputing Applications through allocation CHM250060 from the Advanced Cyberinfrastructure Coordination Ecosystem: Services \& Support (ACCESS) program \cite{access}, which is supported by National Science Foundation grants \#2138259, \#2138286, \#2138307, \#2137603, and \#2138296.

\bibliography{references}

\end{document}


\title{Supplementary material for ``Simulating hydrodynamic interactions in colloidal suspensions using multiparticle collision dynamics with rigid-body constraints''}

\author{Michaela Bush}
\affiliation{Department of Chemical Engineering, Auburn University, Auburn, AL 36849, USA}

\author{Jeremy C. Palmer}
\affiliation{Department of Chemical and Biomolecular Engineering, University of Houston, Houston, TX 77204, USA}

\author{Michael P. Howard}
\email{mphoward@auburn.edu}
\affiliation{Department of Chemical Engineering, Auburn University, Auburn, AL 36849, USA}

\maketitle

\begin{figure}[!ht]
  \centering
  \includegraphics{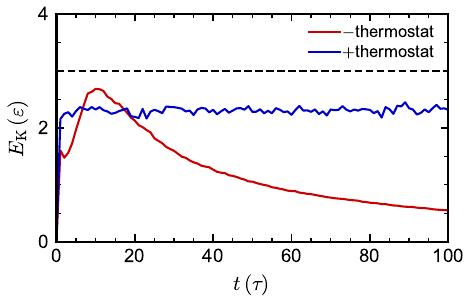}
  \caption{Kinetic energy $E_{\rm K}$ per particle in a bulk hard-sphere colloidal suspension with volume fraction $\phi = 0.10$ using rigid-body constraints without thermalization method. Without ($-$) a thermostat, the energy tends toward zero at long times. With ($+$) a thermostat, the wrong energy is achieved compared to the expected value of $3\,\varepsilon$ based on Eq.~(6).}
\end{figure}

\begin{figure}[!ht]
  \centering
  \includegraphics{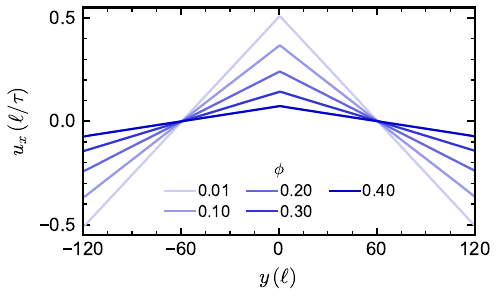}
  \caption{Mass-averaged velocity $u_x$ as a function of position $y$ in reverse nonequilibrium simulations of suspension viscosity at various volume fractions $\phi$ [see Fig.~5(a)].}
\end{figure}